\documentclass{ws-ijmpa}
\usepackage[super,compress]{cite}
\usepackage{graphicx}
\usepackage{epsfig}
\usepackage{color}

\newcommand{\be}{\begin{equation}}
\newcommand{\ee}{\end{equation}}
\newcommand{\bea}{\begin{eqnarray}}
\newcommand{\eea}{\end{eqnarray}}
\newcommand{\beq}{\begin{equation}}
\newcommand{\eeq}{\end{equation}}
\newcommand{\nn}{\nonumber}

\def\ga{\mathrel{\mathpalette\fun >}}
\def\fun#1#2{\lower3.6pt\vbox{\baselineskip0pt\lineskip.9pt
\ialign{$\mathsurround=0pt#1\hfil##\hfil$\crcr#2\crcr\sim\crcr}}}

\begin{document}

\title{HADRON DIFFRACTIVE SCATTERING
AT ULTRAHIGH ENERGIES AND COULOMB INTERACTION}

\author{V.V. ANISOVICH and V.A. NIKONOV}
\address{National Research Centre "Kurchatov Institute",
Petersburg Nuclear Physics Institute, Gatchina 188300, Russia}

\maketitle

\begin{history}
\received{Day Month Year}
\revised{Day Month Year}
\end{history}

\begin{abstract}
We study the interplay of hadronic and Coulomb interactions for $pp$
scattering at LHC energies on the basis of the previous determination
of the real part of the amplitude [{\it V.V. Anisovich, V.A. Nikonov,
J. Nyiri}, Int. J. Mod. Phys. A{\bf 30}, 1550188 (2015)]. The
interference of hadron and Coulomb interactions is discussed in terms
of the $K$-matrix function technique. Supposing the black disk mode for
the asymptotic interaction of hadrons, we calculate interference
effects for the energies right up to $\sqrt{s}= 10^6$ TeV. It turns out
that the real part of the amplitude is concentrated in the impact
parameter space at the border of the black disk that results in a
growth of interplay effects with the energy increase.
  \end{abstract}
\ccode{PACS numbers: 13.85.Lg, 13.85.-t, 13.75.Cs, 14.20.Dh}

\section{Introduction}

At energies of LHC
\cite{Latino:2013ued,Aad:2012pw,Khachatryan:2015gka,Abelev:2012sea} the profile
function of the $pp$-scattering amplitude, $T(b)$, reaches the black
disk limit at small $b$. The black disk picture corresponds to the
non-coherent parton interactions in hadron collisions. For the black
disk scenario the profile function at $\sqrt{s}\ga 100$ TeV gets frozen
inside the disk area, $T(b)\simeq 1$ at $b< R_{black\;disk}$, and the
increasing radius of the black disk, $R_{black\;disk}$, determines the
total, elastic and inelastic cross sections: $\sigma_{tot}\simeq 2\pi
R^2_{black\;disk}$, $\sigma_{el}\simeq \pi R^2_{black\;disk}$ and
$\sigma_{inel}\simeq \pi R^2_{black\;disk}$.

Hadron physics at ultrahigh energies is a physics of large
energy logarithms, $\ln s\equiv \xi>>1$,
\cite{Froissart:1961ux,Gaisser:1988ra,Block:1990hq,Fletcher:1992sy,Dakhno:1999fp}
and increasing parton disks
\cite{Anisovich:2013lba,Anisovich:2013sya,Anisovich:2015geh}.
Considering an interplay of hadron and Coulomb interactions
in this energy region we concentrate our
attention on the black disk mode.

Using notations of ref. \cite{Anisovich:2015geh} we present the
hadronic scattering amplitude with switched off Coulomb interaction
as follows:
\bea \label{1}
&&
A({\bf q}^2 ,\xi)= \int d^2b e^{i{\bf q}{\bf b}}T(b,\xi),
\\
&&
T(b,\xi)=1-\eta(b,\xi)\exp{(2i\delta(b,\xi)})
= \frac{-2i K(b,\xi)}{1-iK(b,\xi)}\,,
\nn
\eea
where $\xi=\ln s$,
$b=|{\bf b}|$. For the profile function we write:
\be \label{e2}
T(b,\xi)=T_{\Im}(b,\xi)-iT_{\Re}(b,\xi),\qquad
  T_{\Re}(b,\ln s)
\simeq\frac{\pi}{2}\frac{\partial T_{\Im}(b,\ln s)}{\partial (\ln s)} \,.
\ee
At the asymptotic regime the imaginary part of the amplitude
is a generating function for $A_{\Re}({\bf q}^2,\ln s)$  that is based on
asymptotic equality $[\sigma_{tot}(pp)/\sigma_{tot}(p\bar
p)]_{\sqrt{s}\to\infty}=1$
  (see \cite{Anisovich:2015geh} for detail). The total and diffractive
cross sections read:
  \be
\sigma_{tot}=2 \int d^2b T_{\Im}(b,\xi),  \qquad
4\pi\frac{d\sigma_{el}}{d{\bf q}^2}=
(1+\rho^2)A_{\Im}^2({\bf q}^2)\,,
\ee
with the usual
notation $A^{H}_{\Re}({\bf q}^2,\xi)/A^{H}_{\Im}({\bf q}^2,\xi)=\rho({\bf q}^2,\xi)$.
Taking into account that $\rho^2$ is small,  $\rho^2\sim 0.01$, one can
approximate:
\be \label{c5}
\Big|A_{\Im}({\bf q}^2 ,\xi)\Big| \simeq
2\pi^\frac12\sqrt{
\frac{d\sigma_{el}}{d{\bf q}^2}
},
\ee
that makes direct calculations of the real part of the scattering amplitude,
$A_{\Re}({\bf q}^2 ,\xi)$, possible, basing on the energy dependence of the
diffractive scattering cross section.
The corresponding calculations were performed in ref.
\cite{Anisovich:2015geh,Anisovich:2015aha}
using the data at $\sqrt{s}\sim 5-50$ TeV
\cite{Latino:2013ued,Aad:2012pw,Khachatryan:2015gka,Abelev:2012sea}
and the results of the previous analyses
\cite{Dakhno:1999fp,Anisovich:2013lba,Anisovich:2013sya}.
In ref. \cite{Anisovich:2015geh,Anisovich:2015aha}
the real parts of the hadronic scattering amplitude,
$A_{\Re}({\bf q}^2,\xi)$, are given for a set of energies,
$\sqrt{s}=1,10,10^2,...,10^6$ TeV, as well as profile functions ($T_\Im(b,\xi)$
and
$T_\Re(b,\xi)$) and K-matrix functions ($K_\Im(b,\xi)$ and $K_\Re(b,\xi)$).

In the present paper,
on the basis of results of ref.\cite{Anisovich:2015geh,Anisovich:2015aha},
  we consider a combined action of the Coulomb and hadronic interactions
for the diffractive scattering region. If the eikonal approach works, the
straightforward way to take into account the interplay of these
interactions is the use of the $K$-matrix
function technique. We present the corresponding formulae (Section 2),
results of numerical calculations for LHC energy (Section 3) and predictions
for the larger energies in the case of the black disk mode for asymptotic
regime (Section 4).

\section {Diffractive scattering amplitude at ultrahigh energy and
Coulomb interaction }

The interplay of hadronic and Coulomb interactions was studied in a set of
papers, see
\cite{Bethe:1958zz,solo,West:1968du,Franco:1973ei,Cahn:1982nr,Kundrat:1993sv,Kaspar:2011zz}
  and references therein. For
ultrahigh energies and small ${\bf q}^2$,  where the eikonal works,
we present the $K$-matrix function technique which allow directly
  to take  account the combined action of
hadronic and Coulomb interactions ($H+C$).
The corresponding
calculations of $K^{H+C}(b,\xi)$ and the profile function
$T^{H+C}(b,\xi)$ are demonstrated.

\subsection{Interplay of hadronic and Coulomb interactions in the $K$-matrix
function technique}

We consider two types of scattering amplitudes
and corresponding profile functions:
the amplitude with combined interaction taken into account,
$A^{C+H}({\bf q}^2 ,\xi)$ and $T^{C+H}(b,\xi)$,
and that with the switched-off Coulomb interaction,
$A({\bf q}^2 ,\xi)$ and $T(b,\xi)$.

For the combined
interaction profile function we write:
\bea \label{6}
  T^{C+H}(b,\xi)=
\frac{-2iK^{C+H}(b,\xi)}{1-iK^{C+H}(b,\xi)}
=
\frac{-2i\left(K^{C}(b)+K(b,\xi)\right)}{1-i\left(K^{C}(b)
+K(b,\xi)\right)}\,,
\eea
where the Coulomb interaction is written as:
\bea
\label{13}
A^C({\bf q}^2 )&=&
\pm i f_1({\bf q}^2)\frac{4\pi\alpha}{{\bf q}^2+\lambda^2}
f_2({\bf q}^2),
\nn \\
-2i K^C(b)&=&
\pm i\int\frac{d^2q}{(2\pi)^2 }  e^{i{\bf q}{\bf b}}
f_1({\bf q}^2)
\frac{4\pi\alpha}{{\bf q}^2+\lambda^2} f_2({\bf q}^2)\,.
\eea
Here  $\alpha =1/137$; the upper/lower signs refer to the same/opposite charges
of the colliding particles.
The cutting parameter $\lambda$, which removes the infrared divergency,
can tend to zero
in the final result for $A^{C+H}({\bf q}^2 ,\xi)$.
Colliding hadron form factors, $f_1({\bf q}^2)$ and $f_2({\bf q}^2)$,
guarantee the
convergence of the integrals at ${\bf q}^2\to \infty$;
for the $pp^\pm$ collisions we use:
\be
f_1({\bf q}^2)=f_2({\bf q}^2)=\frac{1}{(1+\frac{{\bf q}^2}{0.71GeV^2})^2} \,.
\ee
The point which should be emphasized, the Eq. (\ref{6}) gives us the amplitude
imposed by the unitarity condition.

\begin{figure*}[ht]
\centerline{\epsfig{file=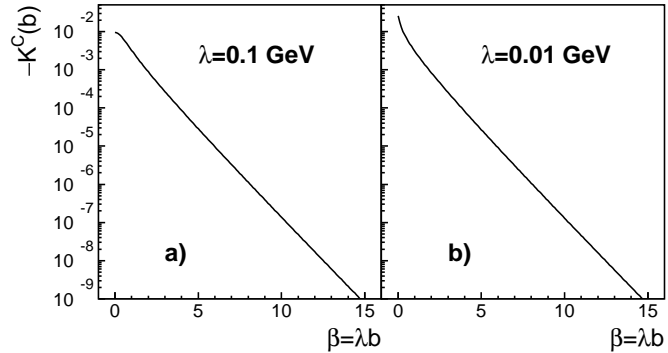,width=0.70\textwidth}}
\caption{The $K$-matrix function, $-K^C(b)$, for the pure Coulomb interaction
in $pp$ collision at different $\lambda$
(a) $\lambda=0.1$ GeV, b) $\lambda=0.01$ GeV): we use
$\beta= \lambda b$ for abscissa. } \label{fig1} \end{figure*}

\begin{figure*}[ht]
\centerline{\epsfig{file=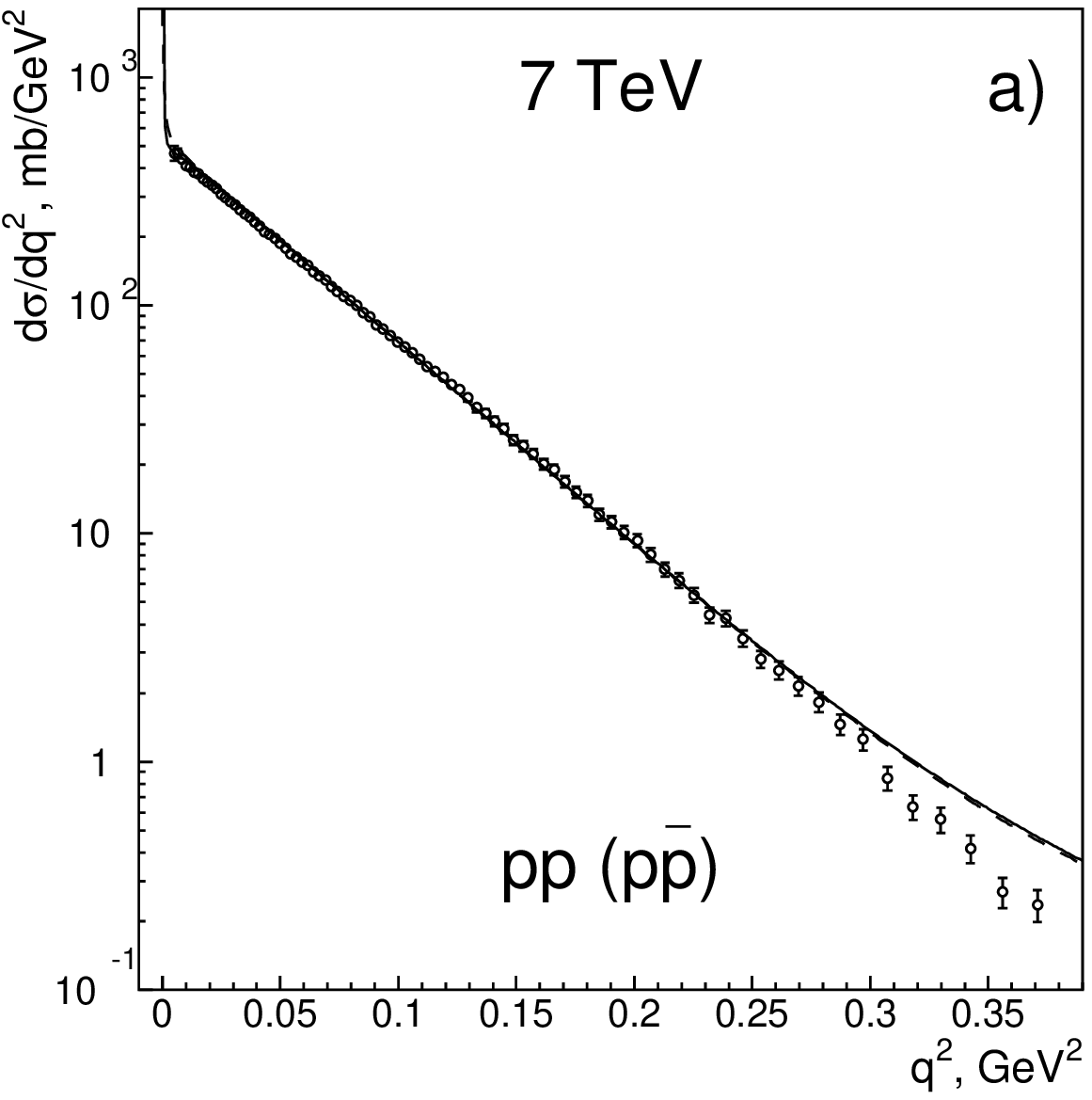,width=0.5\textwidth}
             \epsfig{file=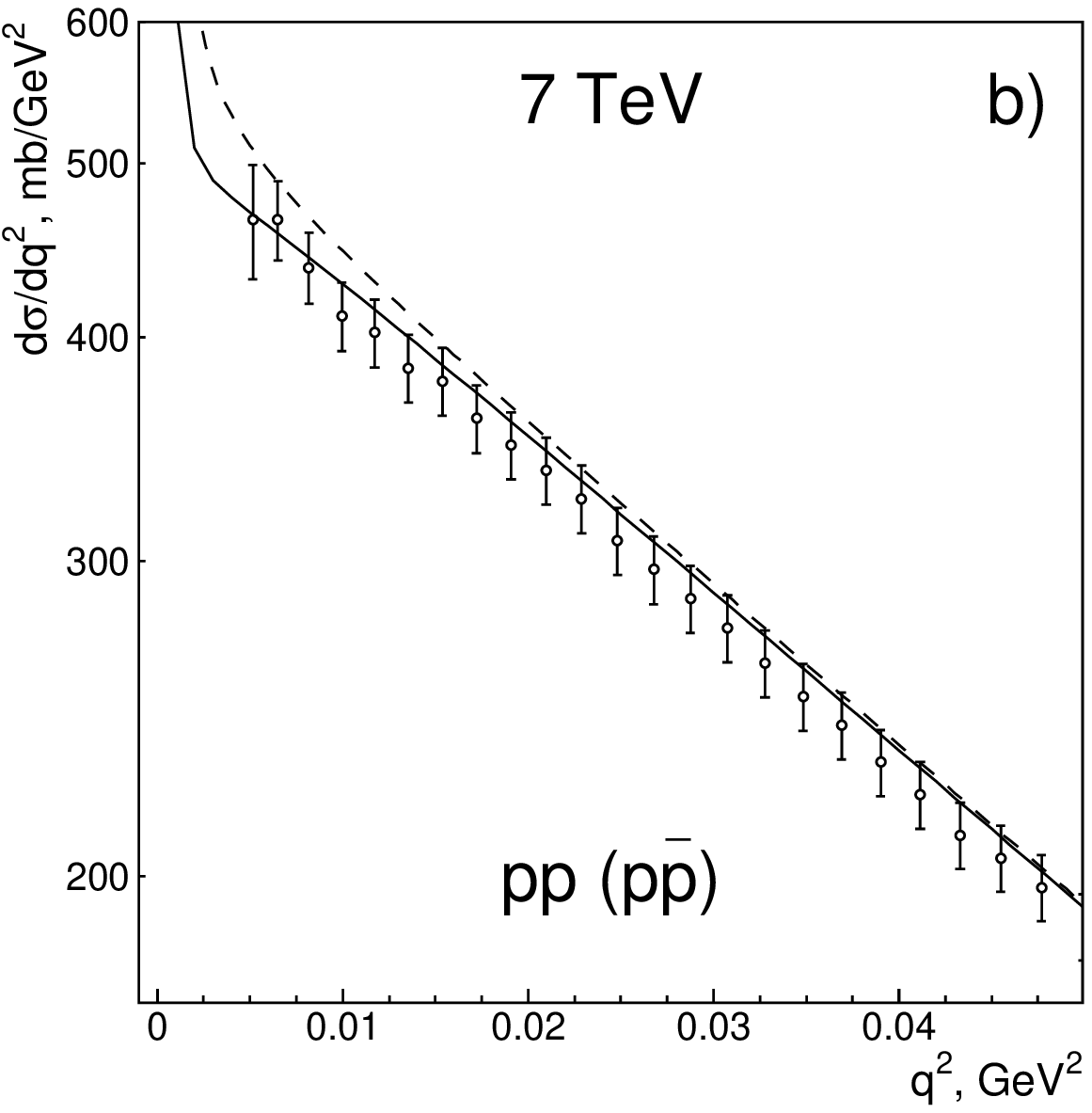,width=0.5\textwidth}}
\centerline{\epsfig{file=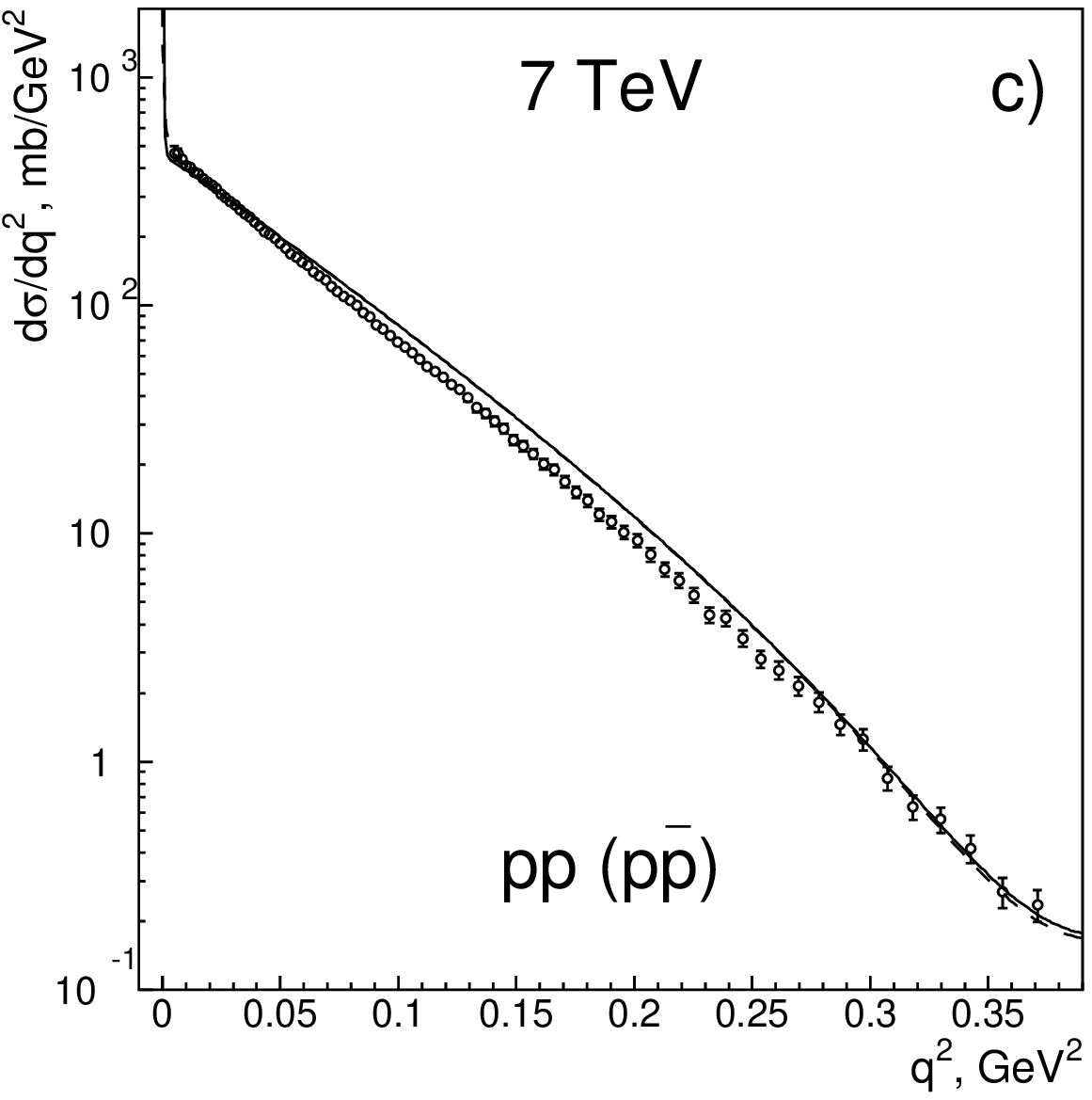,width=0.5\textwidth}
             \epsfig{file=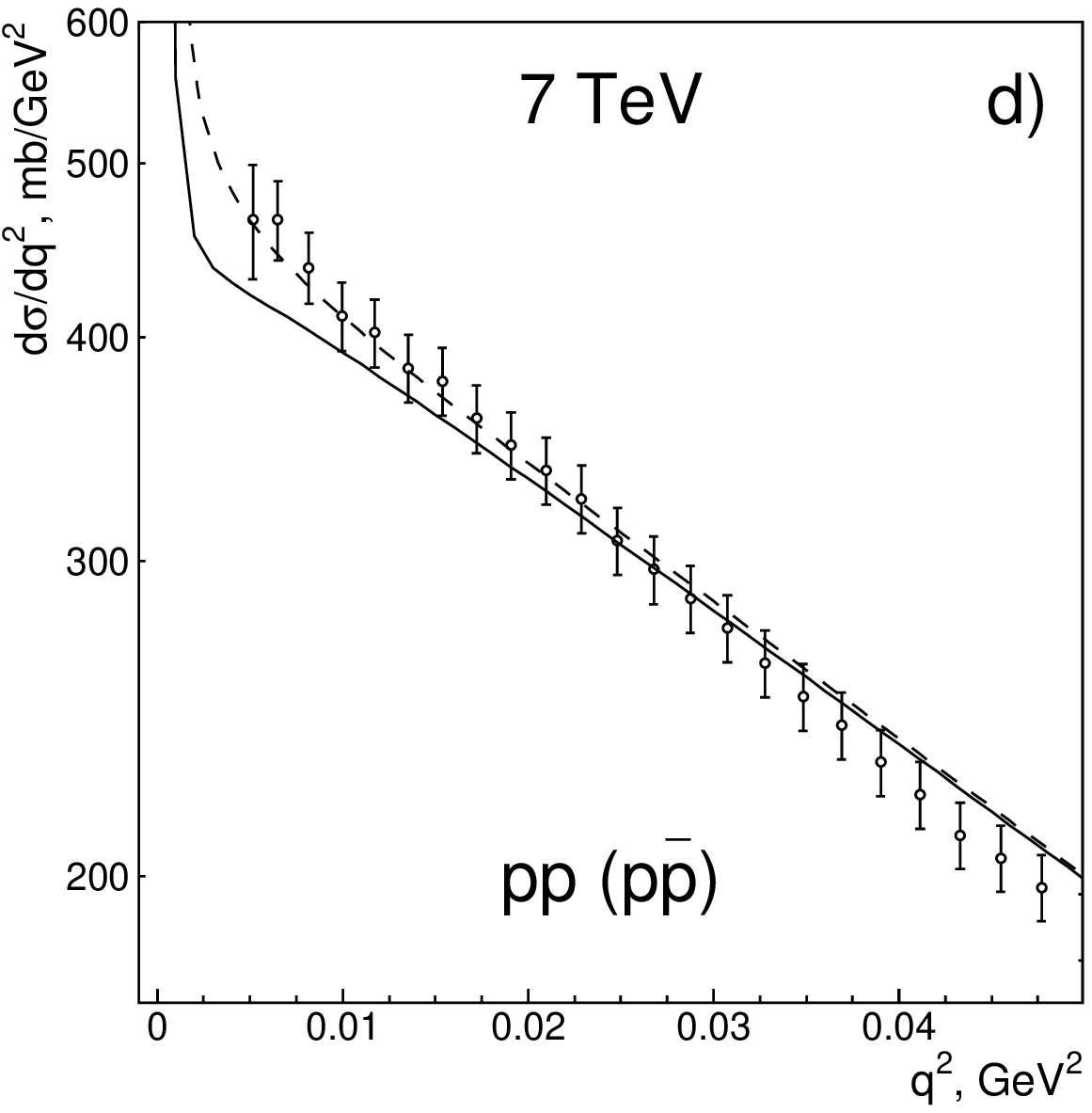,width=0.5\textwidth}}
\caption{
Diffractive scattering cross section for $pp$ at 7 TeV (TOTEM
\cite{Latino:2013ued})
versus description with interplay of hadronic and Coulomb interactions
($\lambda$=0.01 GeV): figures (a,b) refer to vertion (1) for determination of
the hadronic
amplitude, figures (c,d) to version (2); solid curves refer to $pp$, dashed
ones to
$p\bar p$ .} \label{fig2}
\end{figure*}

\section{Diffractive scattering cross section at the LHC energy and
interference of hadronic and Coulomb interactions }

On Fig. {\ref{fig1}} we show $-K^{C}(b) $
for $\lambda =0.1$ GeV (a) and $0.01$ GeV (b).
With these $\lambda $'s we calculate at $\sqrt s=7$ TeV the profile function
$T^{C+H}(b,\xi_{LHC})$ and
the corresponding amplitude $A^{C+H}({\bf q}^2 ,\xi_{LHC})$.  Determination of
the hadronic amplitude, $A_\Im({\bf q}^2 ,\xi_{LHC})$, is performed in
terms of two versions:\\
1) with a direct application of the approximation (\ref{c5}) to the TOTEM
data \cite{Latino:2013ued},\\
2) Using the the results of the Dakhno-Nikonov model
\cite{Dakhno:1999fp,Anisovich:2013lba}.

The description of the data for $\frac{d\sigma_{el}({\bf q}^2
,\xi_{LHC})}{d{\bf q}^2}$ in terms of these two
versions is shown in Fig. \ref{fig2}: here Figs. \ref{fig2}a,b refer to
the version (1) and Figs. \ref{fig2}c,d correspond to version (2).
Let us emphasise that the real part of the hadronic amplitude,
given by Eq. (\ref{e2}), is taken here into account.

The Dakhno-Nikonov model gives a somewhat
worse description of the $\frac{d\sigma_{el}({\bf q}^2 ,\xi_{LHC})}{d{\bf
q}^2}$
at 7 TeV than that using Eq. (\ref{c5}). This is not surprising because
the model  descibes the data
in a broad energy interval, 0.5-50 TeV \cite{Anisovich:2013lba}, and the model
parameters
are responsible for the complete set of the data.

\begin{figure*}[ht]
\centerline{\epsfig{file=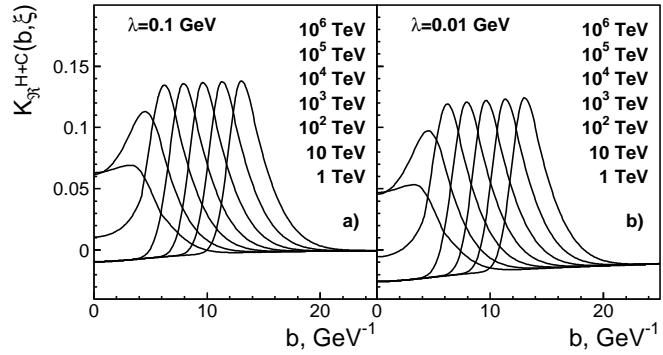,width=0.70\textwidth}
}
\caption{
  Combined hadronic and Coulomb interactions for the $pp$ scattering:
  Real parts of the  $K$-matrix functions at different $\lambda$
(a) $\lambda=0.1$ GeV, b) $\lambda=0.01$ GeV)
 at a set of energies $\sqrt{s}=1,10,10^2,...,10^6$ TeV; at
    $\sqrt{s}\geq 10^2$ TeV the black disk mode is supposed.
}
\label{fig3}
\end{figure*}

\begin{figure*}[ht]
\centerline{\epsfig{file=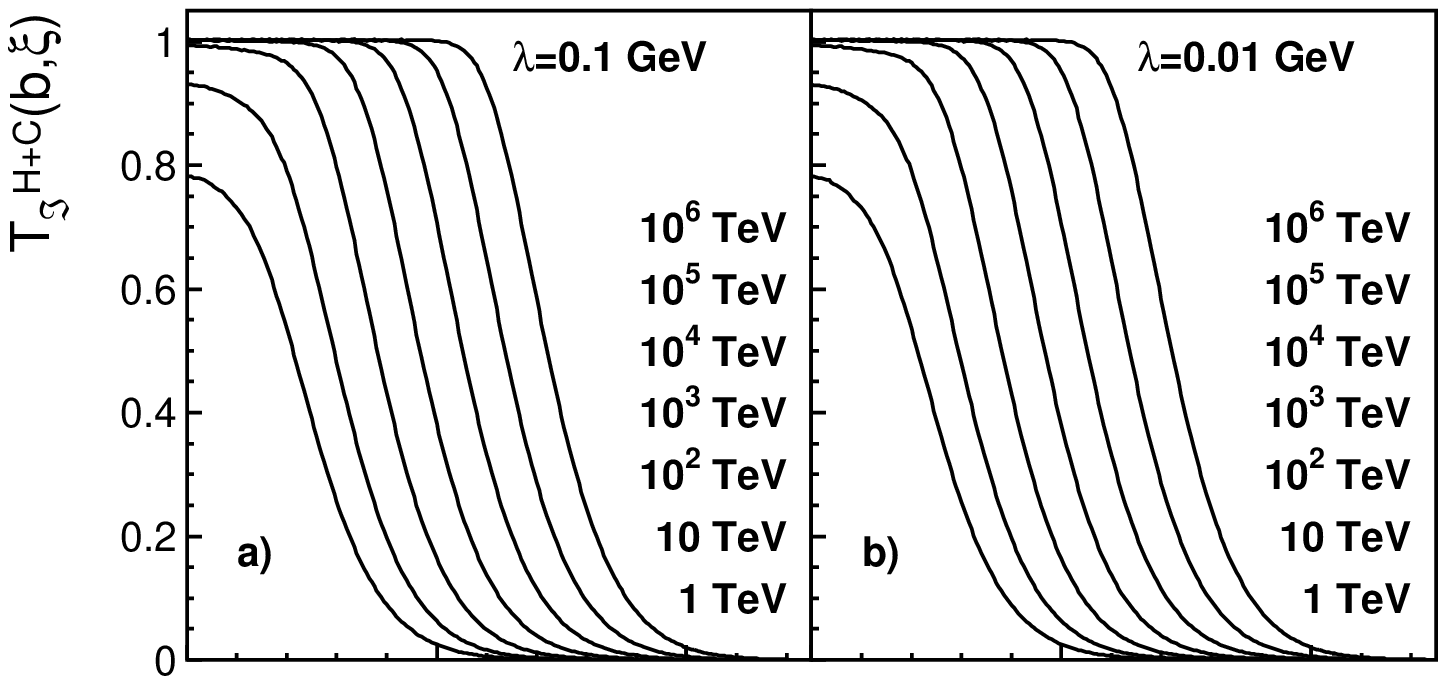,width=0.70\textwidth}}
\vspace{-8.5mm}
\centerline{\epsfig{file=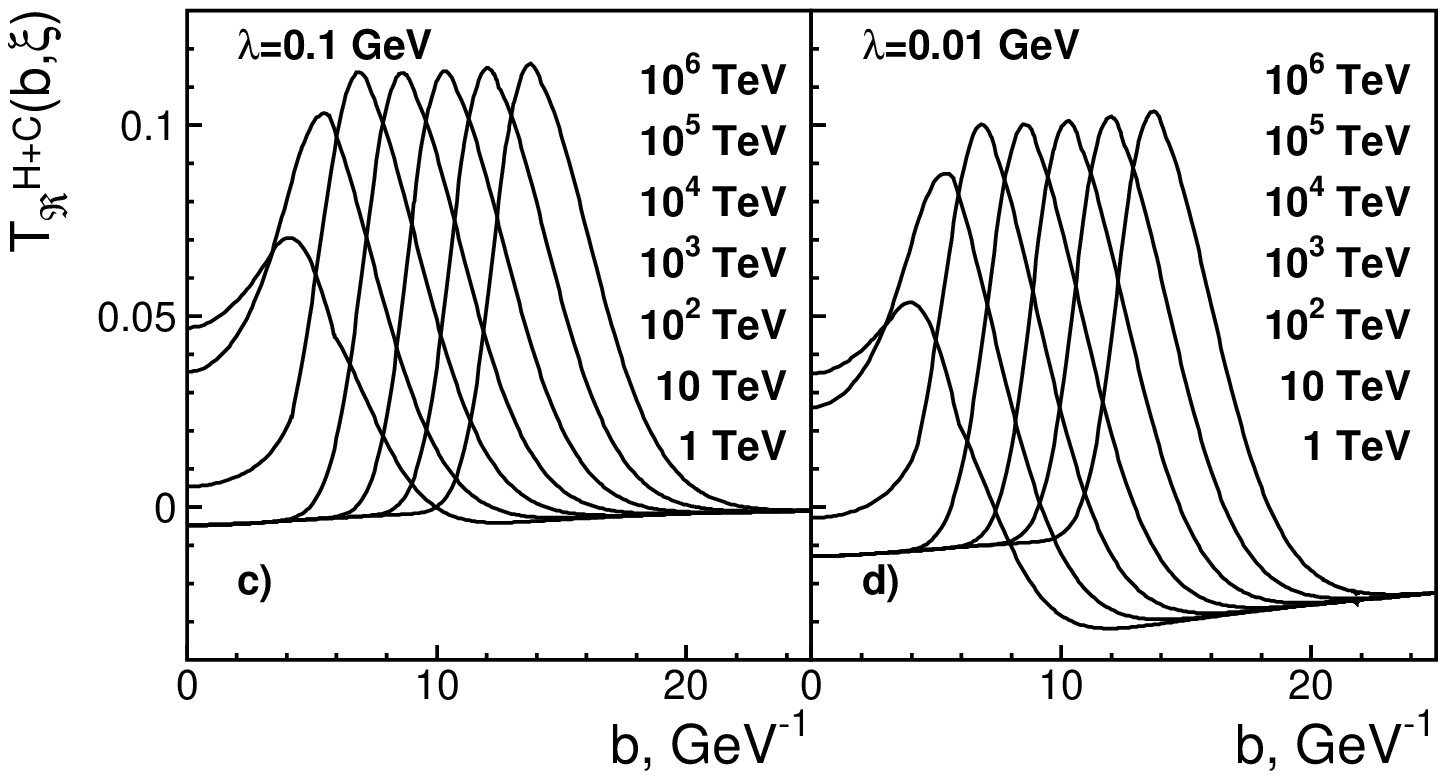,width=0.70\textwidth}}
\caption{
  Combined hadronic and Coulomb interactions for the $pp$ scattering:
  Imaginary (a,b) and real parts (c,d) of the  profile function at
  different $\lambda$
(a,c) $\lambda=0.1$ GeV, b,d) $\lambda=0.01$ GeV)
for a set of energies $\sqrt{s}=1,10,10^2,...,10^6$ TeV; at
    $\sqrt{s}\geq 10^2$ TeV the black disk mode is supposed.
}
\label{fig4}
\end{figure*}

\section{Black disk mode: predictions for ultrahigh energies}

  The inclusion of the Coulomb interaction into consideration of hadron
diffractive
scattering does not change the imaginary part of the $K$-matrix function,
  $K_\Im^{H+C}(b,\xi)=K_\Im(b,\xi)$. The real part of the $K$-matrix
  function for $pp$ scattering, $K_\Re^{H+C}(b,\xi)=K_\Re^{}(b,\xi)+K^{C}(b)$,
is shown in
  Fig. {\ref{fig3}} for $b<25$ GeV$^{-1}$.

Imaginary and real parts of the profile functions, $T^{H+C}(b ,\xi)$
for $b<25$ GeV$^{-1}$, are shown in Fig. {\ref{fig4}}; the inclusion of
the Coulomb interaction leads to
considerable perturbations in the real part, Figs. {\ref{fig4}c,d}.

\begin{figure*}[ht]
\centerline{\epsfig{file=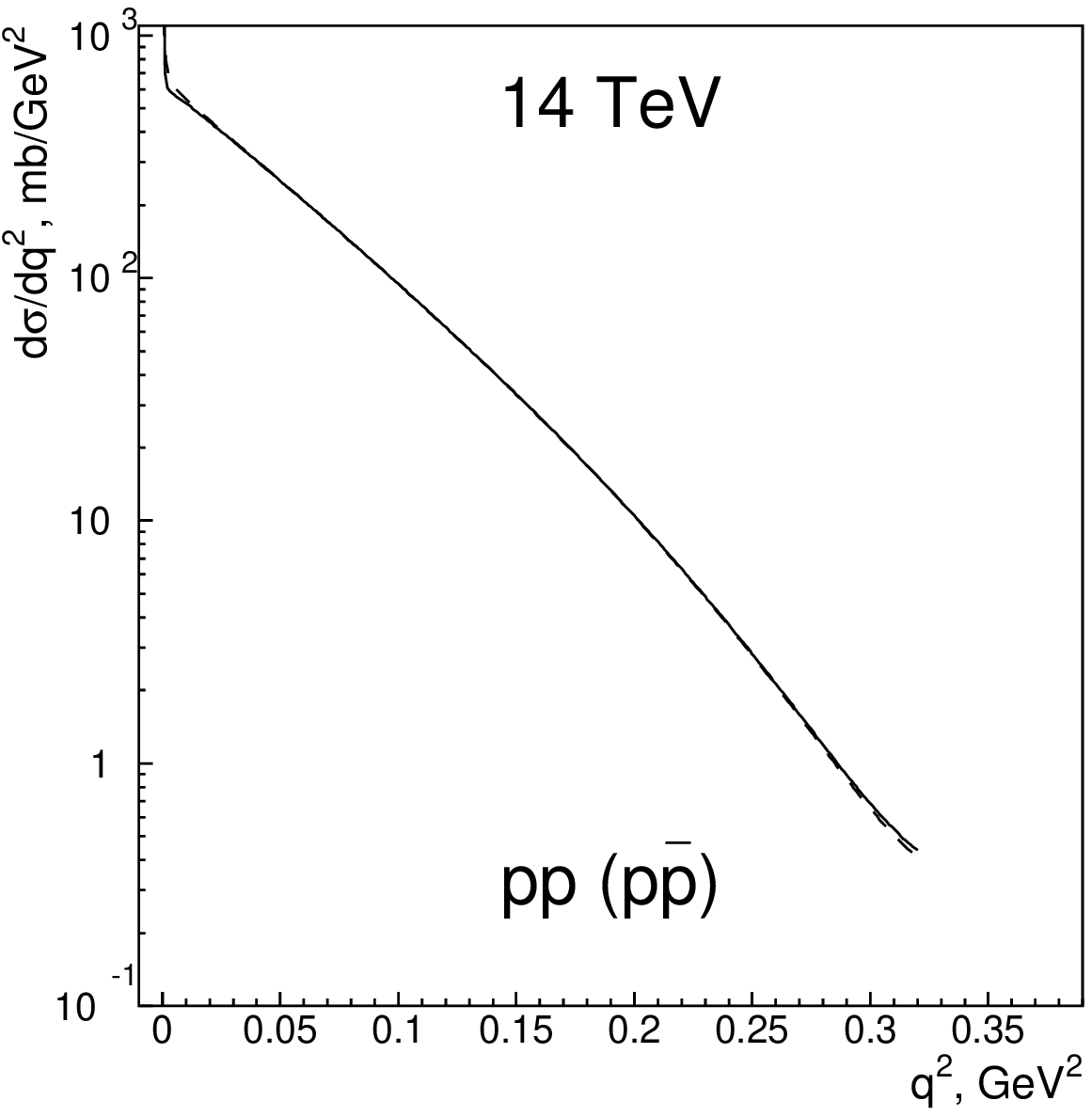,width=0.5\textwidth}
             \epsfig{file=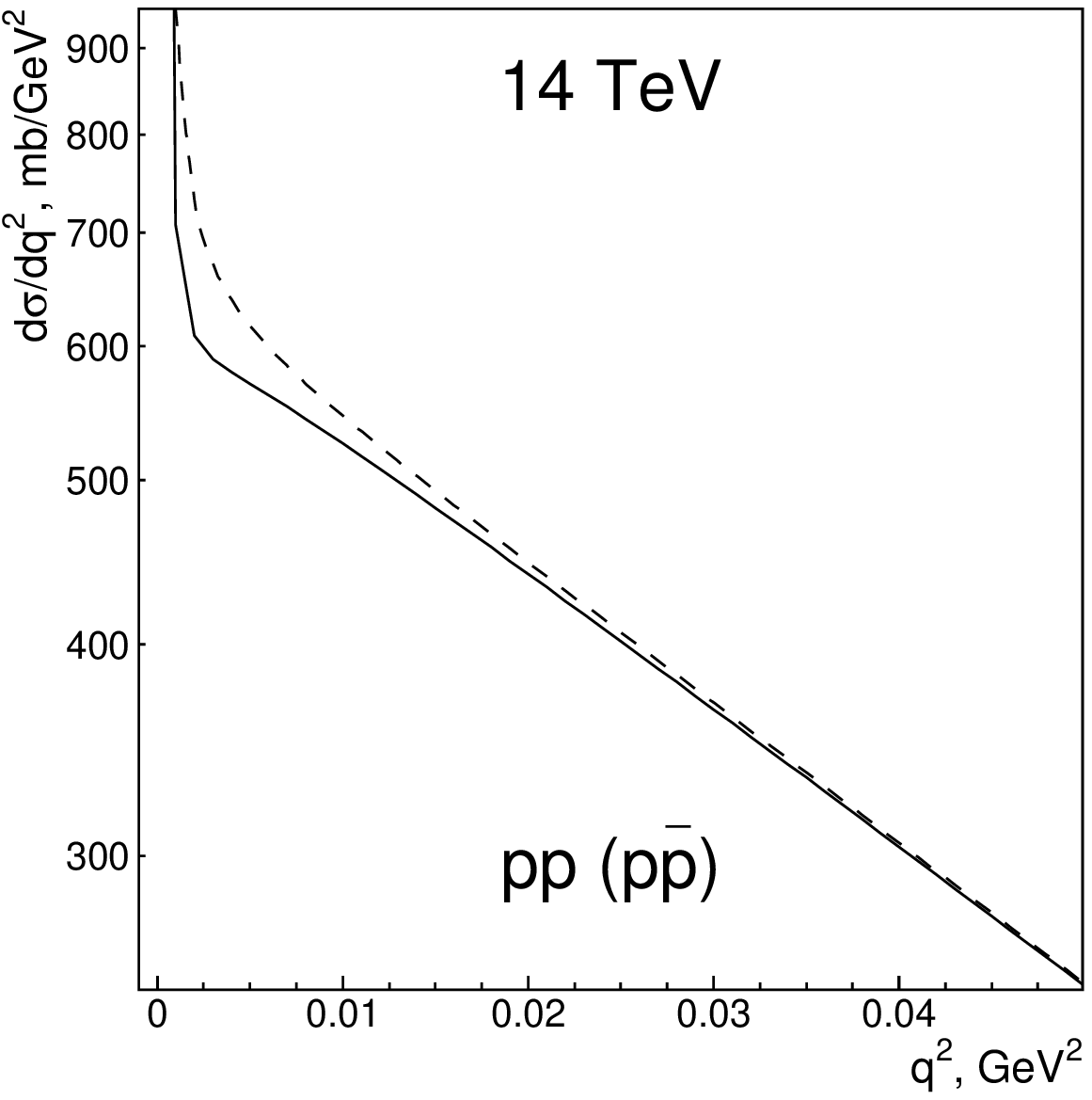,width=0.5\textwidth}}
\caption{
Diffractive scattering cross sections for $pp^\pm$ at 14 TeV.
  The real part of the hadronic amplitude as well as the
interplay of the Coulomb and hadronic interactions
are taken into account
  ($\lambda$=0.01 GeV):
the solid curves refer to $pp$, dashed ones to
$p\bar p$ .}
\label{fig5}
\end{figure*}

The predicted diffractive scattering cross sections
$\frac{d\sigma_{el}({\bf q}^2)}{d{\bf q}^2}$ for $pp^\pm$ at 14 TeV are shown
in Fig.  \ref{fig5}. Recall that we take into account
here the real part of the hadronic amplitude as well as the interplay of
the Coulomb and hadronic interactions.

\section{Conclusion}

The interplay of the hadronic and Coulomb
interactions at very small ${\bf q_\perp}^2$ is discussed in terms of the
$K$-matrix
function. A specificity of the scattering amplitude at ultrahigh energy is
dominance of
the mass-on-shell contributions in intermediate rescattering states that
results in the
mass-on-shell origin of the $K$-matrix functions. Such $K$-matrix functions
allow
to incorporate the Coulomb interaction terms into the scattering amplitude
straightforwardly by using a determination consistent with unitarity
condition, $K^C(b)= \tan\delta^C(s,b)$.
We present the corresponding formulae and perform calculations of
$\frac{d\sigma_{el}({\bf q}^2)}{d{\bf q}^2}$ for $pp^\pm$ at 7 TeV
(Fig. \ref{fig2}) and 14 TeV (Fig. \ref{fig5}).

For ultrahigh energies we calculate  $K^{H+C}(b ,\xi)$ and $T^{H+C}(b ,\xi)$
supposing the black disk mode. However,
the asymptotic high energy regime for diffractive hadron scatterings is not
determined
yet and the resonant disk regime
\cite{Troshin:2014rva,Dremin:2014eva,Anisovich:2014wha} is not
excluded at $\sqrt s > 10^2$ TeV. Therefore an immediate task is to study
the interplay of
the Coulomb and hadronic  interactions in the resonant disk mode.

\subsubsection*{Acknowledgment}

We thank Y.I. Azimov, J. Nyiri and A.V. Sarantsev for useful discussions.
  The work was supported by the Russian Science Foundation grant.

\end{document}